\documentclass[journal=jaccck,manuscript=article]{achemso}
\setkeys{acs}{articletitle = true}
\usepackage{chemformula} % Formula subscripts using \ch{}
\usepackage[T1]{fontenc} % Use modern font encodings

\author{Krishna Prasad Maity}
\email{makrishna@iisc.ac.in}
\affiliation{Department of Physics, Indian Institute of Science, Bangalore 560012, India}
\alsoaffiliation{Université de Strasbourg, CNRS, Institut de Physique et Chimie des Matériaux de Strasbourg, UMR 7504, 23 rue du Loess, 67000 Strasbourg, France}
\author{Ananya Patra}
\affiliation{Department of Physics, Indian Institute of Science, Bangalore 560012, India}
\alsoaffiliation{School of Physics and Astronomy, Tel Aviv University, 6997801 Tel Aviv, Israel}
\author{Narendra Tanty}
\affiliation{Department of Physics, Indian Institute of Science, Bangalore 560012, India}
\author{V Prasad }
\affiliation{Department of Physics, Indian Institute of Science, Bangalore 560012, India}
\title[An \textsf{achemso} demo]
  {Magnetic field driven dielectric relaxation in non-magnetic composite medium: a low temperature study}

\begin{document}


\begin{abstract}
	The frequency dependence of dielectric constant for composites of polyaniline (PANI) and multi-walled carbon nanotube (MWCNT) with different degree of functionalization is studied at low temperature (down to 4.2 K) and magnetic field (up to 3 Tesla) applied both in parallel and perpendicular direction of ac electric field. A relaxation phenomenon is observed in all the MWCNT/PANI composites by applying magnetic field in both the directions, below 10$^3$ Hz. However, PANI does not show any relaxation peak with applied magnetic field in either direction. The relaxation peak frequency does not depend on the strength of magnetic field but it varies with temperature and degree of functionalization of MWCNT in composites. This relaxation phenomenon occurs due to the inhomogeneity of the medium of two highly mismatched conductive materials at low temperatures. We have tried to explain our results in the light of Parish and Littlewood theory about magnetocapacitance in nonmagnetic composite.
\end{abstract}

\section{Introduction}
The coupling between magnetization and dielectric constant is a very well-known phenomenon in multiferroics which have potential application in magnetic field sensors, multi-state memory devices~\cite{appmodferro} and so on. This coupling is manifested by several ways in different materials. For instance, BaTiO$_3$-CoFe$_2$O$_3$ composite exhibits the magnetoelectric coupling by the strong elastic interaction between these two phases~\cite{Zheng661}. The external magnetic field affects the magnetisation, and thus influences the dielectric properties of these materials, known as magnetodielectric (MD) effect. However, MD effect can be observed without the coupling between these two. The interfacial polarization~\cite{Catalan} at the grain boundaries can also induce the MD effect. There are few reports where the interfacial polarization and magnetic field give the dielectric relaxation~\cite{Maglione_2008}. For example, La$_{2/3}$Ca$_{1/3}$MnO$_3$ shows the magnetic field dependence dielectric constant above ferromagnetic to paramagnetic transition temperature, which is manifested by the interfacial polarization~\cite{doi:10.1063/1.2213513}. Anomalous dielectric relaxation is observed in doped silicon by applying magnetic field perpendicular to the ac electric field~\cite{PhysRevB.78.045205Brook}. This effect is differentiated from usual Debye like dielectric relaxation which occurs due to parallel magnetic field and is explained by the effect of Lorentz force on charge polarization. Parish and Littlewood have proposed that dielectric relaxation can be observed in inhomogeneous medium without magnetism, where the inhomogeneity gives rise to magnetic field dependent dielectric relaxation due to the mixing between real and imaginary part of the dielectric constant~\cite{PhysRevLett.101.166602Parish}. The interfacial polarization driven MD effect in the composite of graphene with polyvinyl-alcohol (PVA) is explained by this theory~\cite{doi:10.1021/jp203724fMitra}. The MD effect is also observed in composites of reduced graphene oxide (RGO) with different polymers, which occurs due to interfacial polarization, and indicates the interaction between RGO and polymer~\cite{Pradhan_2020}. 
\\ In this article, we have studied the dielectric properties of PANI and composite with different degree of functionalized MWCNT at low temperature (down to 4.2 K) applying magnetic field up to 3 T. This composite is an inhomogeneous system and becomes mixture of highly mismatched conductive components at low temperature. The degree of functionalization of MWCNT changes the interfacial polarization between fMWCNT and PANI. Therefore, this system is very suitable to study the interfacial polarization in the presence of external magnetic field. This study will serve to understand the dielectric relaxation of non-magnetic inhomogenious medium with applied magnetic field.

\section{Results and discussion}

%\subsection{Outline}

\begin{figure}[tbh!]
	\begin{center}
		\includegraphics[width=0.8\textwidth]{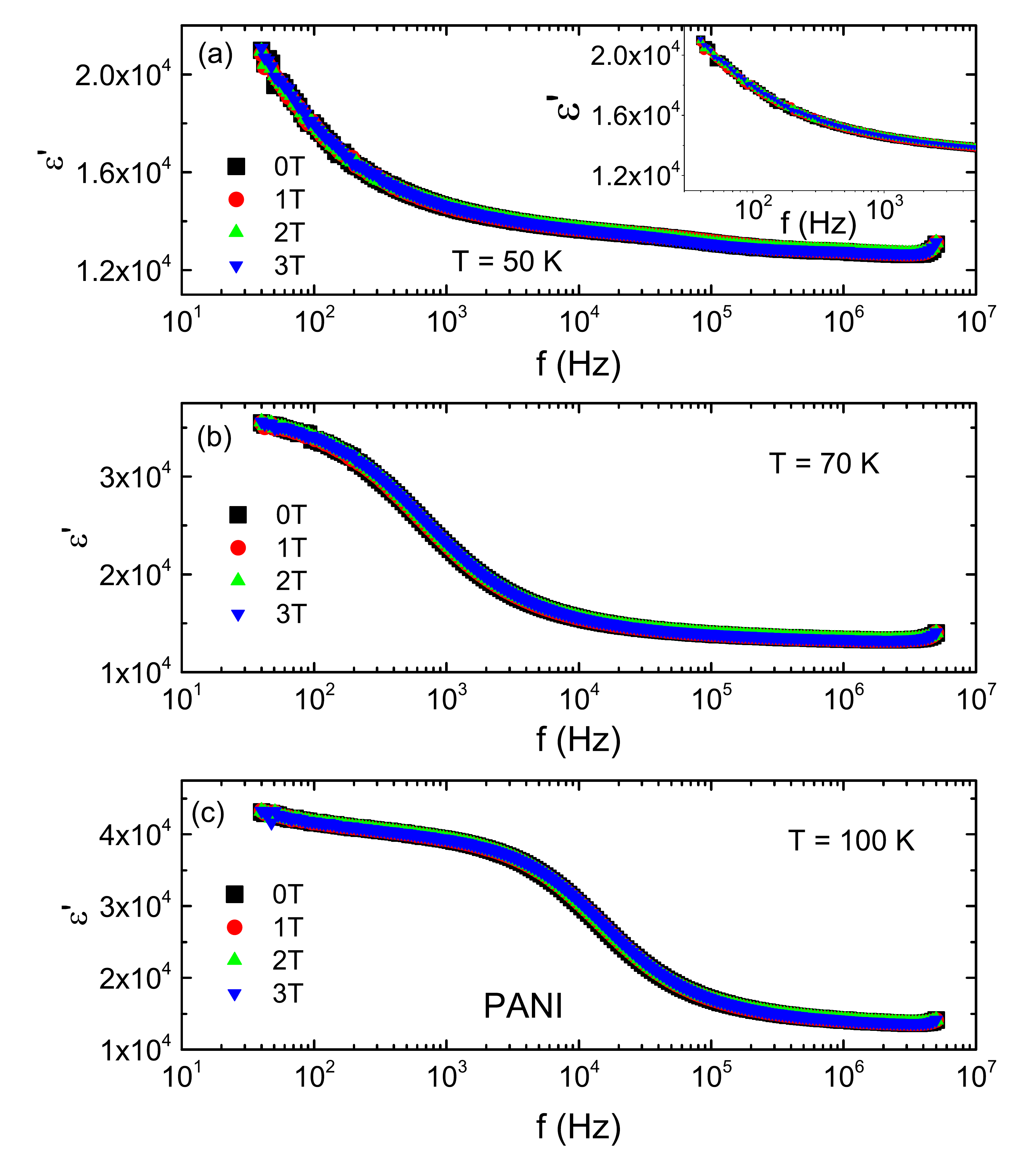}
		\small{\caption{ Variation of $\epsilon^\prime$ with frequency for PANI by applying magnetic field at temperatures (a) 50 K, (b) 70 K and (c) 100 K. No relaxation peak is observed at low frequency due to the applied magnetic fields. [Inset graph represents the $\epsilon^\prime$ variation with magnetic field, zoomed in the frequency range 40 Hz - 5 kHz at T = 50 K].  \label{fig: pani}}}
	\end{center}
\end{figure}

\begin{figure}[tbh!]
	\begin{center}
		\includegraphics[width=0.8\textwidth]{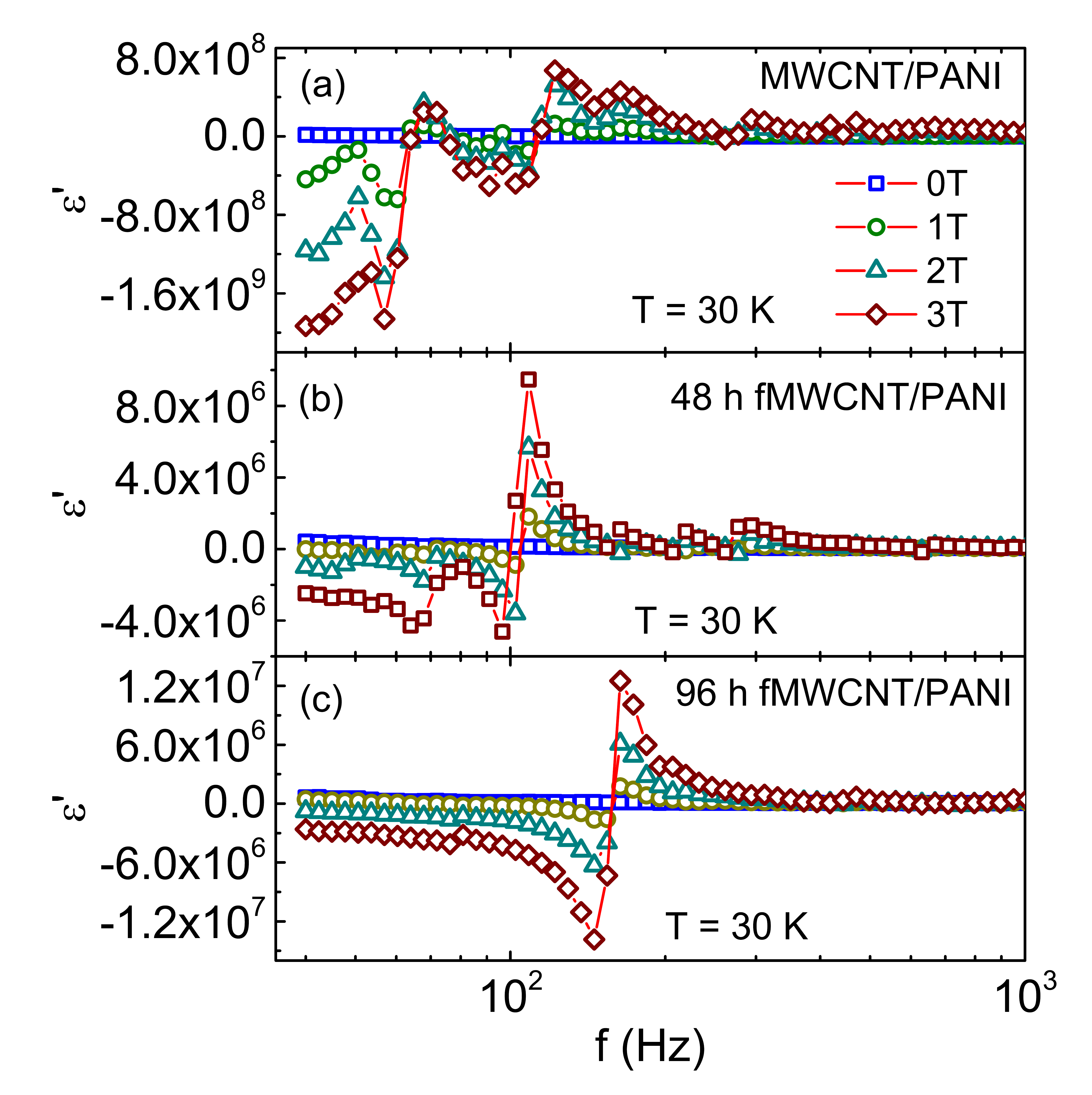}
		\small{\caption{Variation of real part of dielectric constant with frequency applying magnetic field at temperature 30 K for (a) MWCNT/PANI (0 h), (b) 48 h and (c) 96 h fMWCNT composite with PANI. The relaxation shifts towards higher frequency with increasing degree of functionalization of MWCNT. [The legend of graph (a) is also shared by (b) and (c).] \label{fig: 30k}}}
	\end{center}
\end{figure}

\begin{figure}[tbh!]
	\begin{center}
		\includegraphics[width=0.8\textwidth]{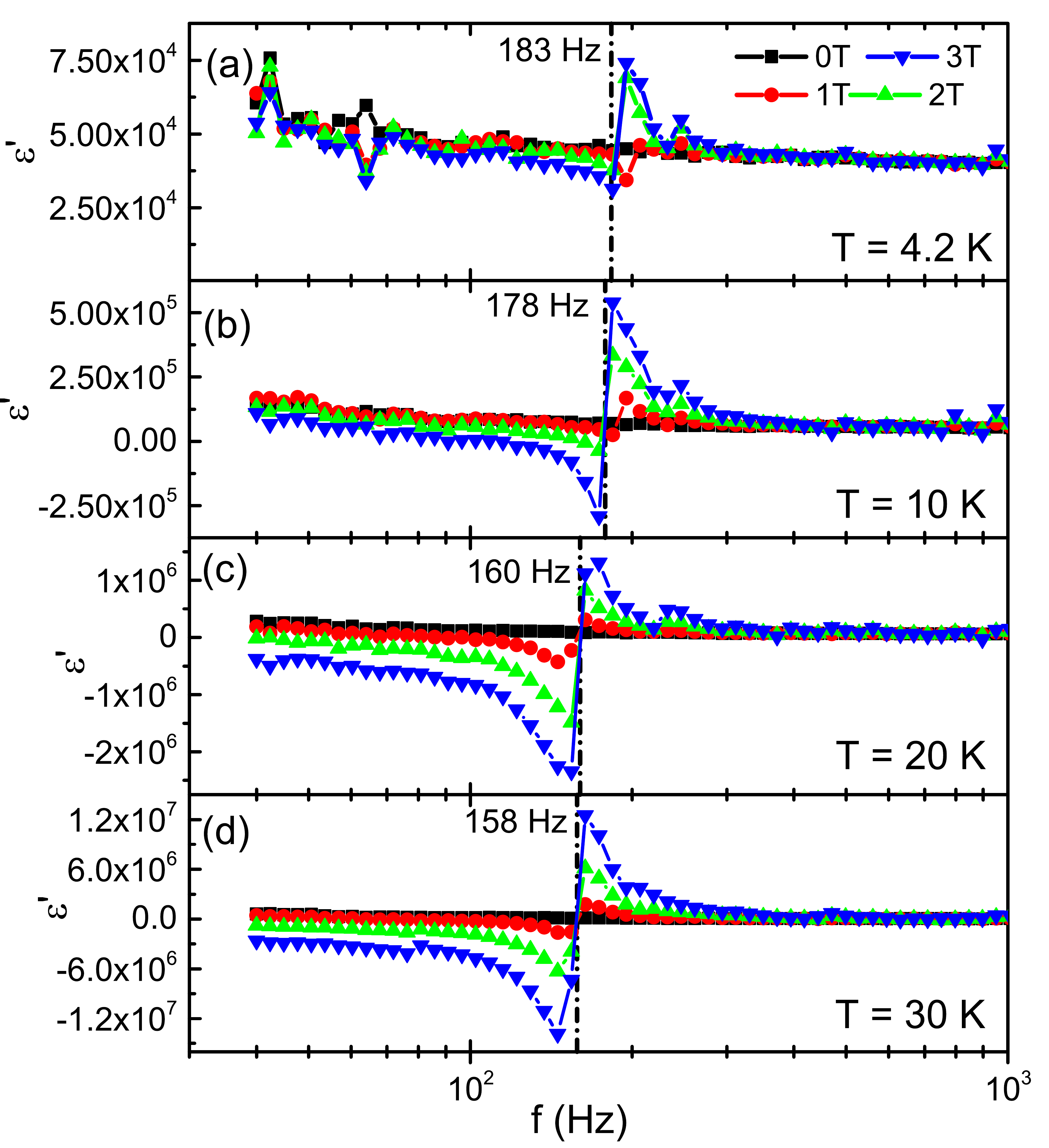}
		\small{\caption{ Variation of real part of dielectric constant with frequency for 96 h fMWCNT/PANI composite applying magnetic field at different temperature (a) 4.2 K, (b) 10 K, (c) 20 K and (d) 30 K. The relaxation shifts towards lower frequency with increasing temperature.  \label{fig: 96h}}}
	\end{center}
\end{figure}

\begin{figure}[tbh!] 
	\begin{center}
		\includegraphics[width=0.8\textwidth]{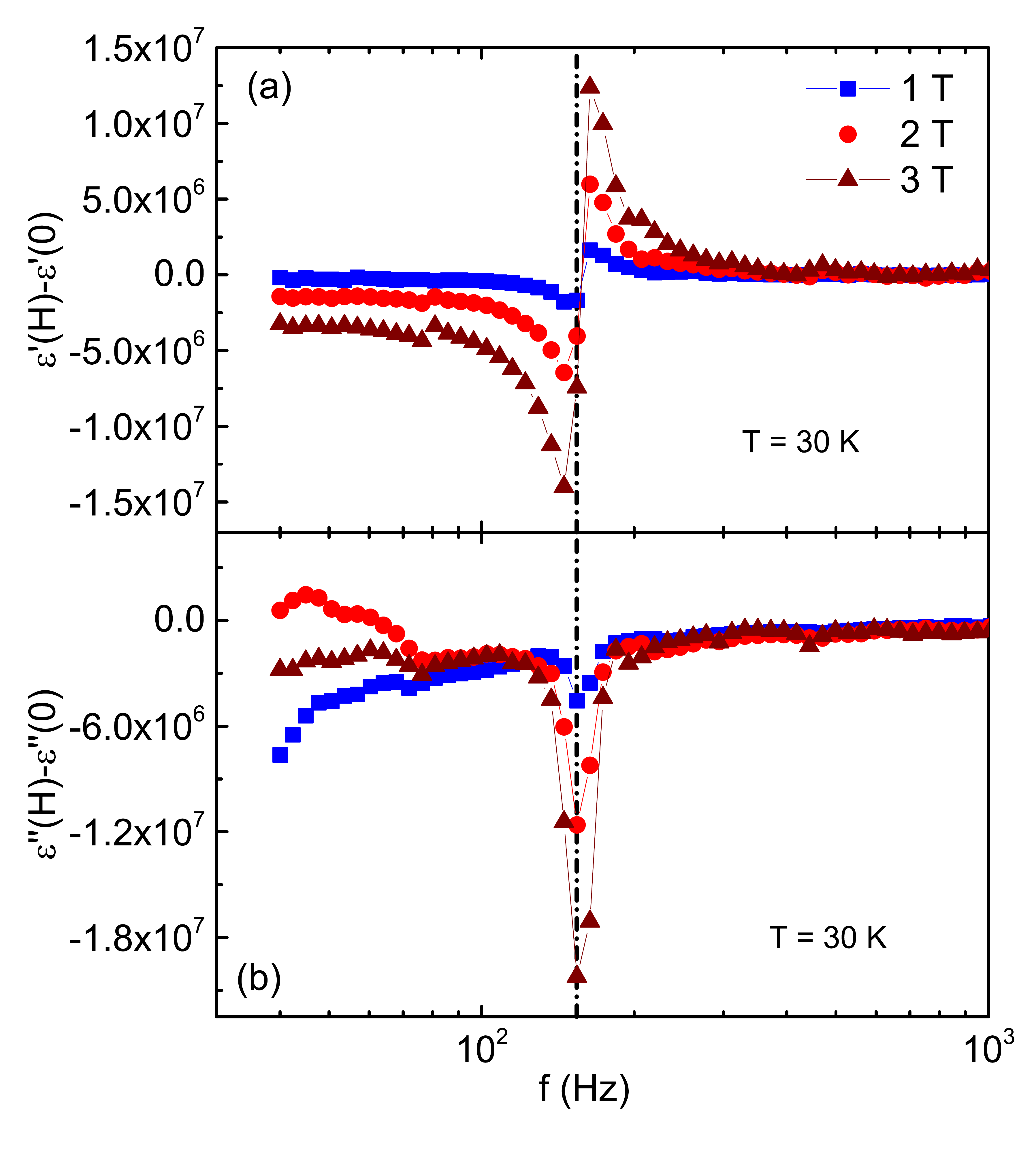}
		\small{\caption{ The difference between with and without applying magnetic field of (a) real and (b) imaginary parts of dielectric constant as a function of frequency for 96 h fMWCNT/PANI composites at 30 K.  \label{fig: 96h_30K}}}
	\end{center}
\end{figure}

\begin{figure}[tbh!]
	\begin{center}
		\includegraphics[width=0.8\textwidth]{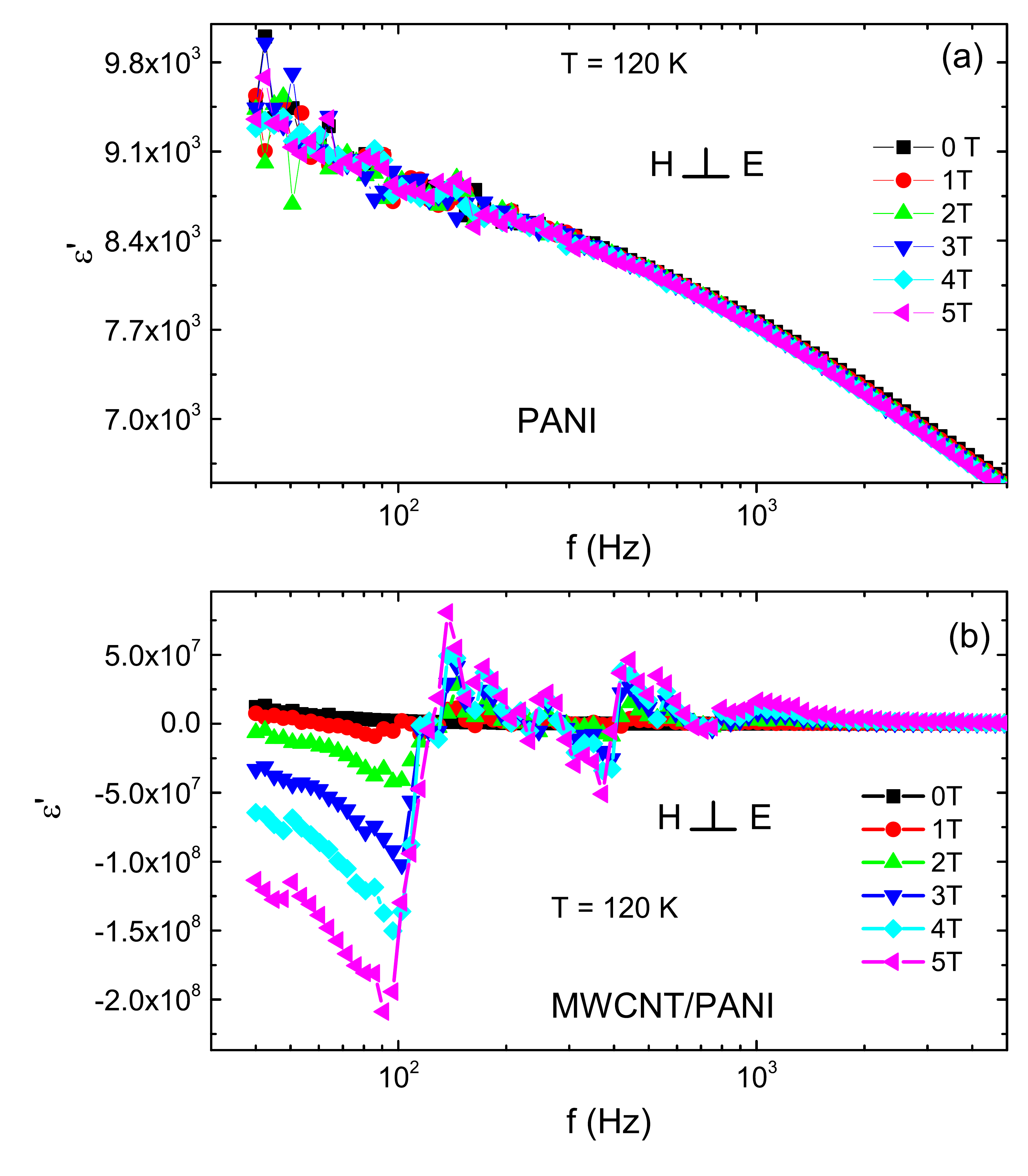}
		\small{\caption{ Variation of real part of dielectric constant with frequency for (a) PANI and (b) MWCNT/PANI applying magnetic field perpendicular to the ac electric field.  \label{fig: ac2}}}
	\end{center}
\end{figure}

In Fig.~\ref{fig: pani}, the variation of real part of dielectric constant ($\epsilon^\prime$) with frequency is shown for PANI at different temperatures by applying magnetic field parallel to the current. We can observe that $\epsilon^\prime$ decreases with increasing frequency and is constant at high frequency. The decreasing behaviour represents the dipolar relaxation in the system. The dipolar relaxation shift towards higher frequency with increasing temperature. All the curves for different applied magnetic field merge with one another at a constant temperature indicating there is no effect of magnetic field in dielectric properties and the electrical polarization behaviour of PANI.
\\ The variation of $\epsilon^\prime$ with frequency is shown in Fig.~\ref{fig: 30k} for different degree of functionalized fMWCNT/PANI composites by applying magnetic field in the parallel direction of ac electric field at temperature 30 K. We can clearly observe that a relaxation in $\epsilon^\prime$ appears due to the magnetic field in all the composites at low frequency ($<$ 1000 Hz). However, this relaxation behaviour is not present in the absence of magnetic field (H = 0 T). This behaviour is completely dissimilar to the PANI. The relaxation frequency does not vary with increasing magnetic field for composites at constant temperature, however it shifts towards higher frequency with increasing the degree of functionalization of MWNT in different composites. The relaxation peak frequency increases from 63 Hz to 160 Hz when the duration of functionalization increases from 0 h to 96 h. The functionalization of MWCNT enhances the interfacial polarization and increases the boundness of charge carrier between fMWCNT and PANI~\cite{10.1088/1361-6463/abd2eb}. The enhanced boundness of charge carriers help to relax fast with applying magnetic field and relaxation frequency shifts towards higher value. This coupling of dielectric constant and magnetic field is not due to the presenece of magnetic relaxation of either PANI or fMWCNT,since PANI manifests the paramagnetism~\cite{NOVAK20101725} and MWCNT shows diamagnetic behaviour at low temperature~\cite{ELLIS2006378}. In our case, the interfacial polarization between fMWCNT and PANI is affected by the magnetic field and give rise to the dielectric relaxation at low frequency.  
\\This magnetic field relaxation behaviour in composites can be explained by Parish and Littlewood theory. At low temperature, PANI becomes highly resistive but MWCNT is very conducting (conductivity difference $\approx$ 10$^3$ - 10$^4$ S/cm). The mismatch of conductivity of two dissimilar dielectric medium creates interfacial polarization which is affected by the magnetic field.  
PANI contains relatively high and low conductive regions depending on the doping. The charge transport between these high and low conductive regions creates the electrical polarization and shows the finite dielectric constant. At low temperature, PANI behaves like a homogeneous system of almost similar conductive regions. In this case, external magnetic field does not affect the dielectric constant and no relaxation is observed at low frequency.	 
At high temperature the  electrode polarization might contibute to the dielectric relaxation phenomenon and induce magnetic field dependence. For proper comparison, we have used same silver plate electrodes for pure PANI and the composites. However, magnetic field dependence is not observed in pure PANI. Hence, contribution from electrode towards magnetic field dependence relaxation can be ruled out.
%In Fig.~\ref{fig: 30k}, the variation $\epsilon^\prime$ with frequency for different applied magnetic field is shown at 30 K. In contrast to PANI,the relaxation peak is observed for all the fCNT/PANI composite at low frequency. This relaxation pear appear due to the inhomogeneous system of two diffrently conducting materials. The position of relaxation peak frequency shifts towards higher value with increasing degree of chemical functionalization. The increasing degree of functionalizaion of CNT inceases the interfacial polarization (Maxwell-Wagner-Sillars). 
%And also, the polar interaction between fCNT and PANI increases, which enhances the boundness of the charge carriers.
\\The variation of $\epsilon^\prime$ with frequency of 96 h fMWCNT/PANI composite is shown in Fig.~\ref{fig: 96h} for different temperatures. The relaxation behaviour is observed at all the temperatures and relaxation frequency shifts towards lower frequency with increasing temperature. The thermal energy of charge carrier increases when temperature increases and deflect longer distance due to enhanced Lorentz force, and charge carriers take longer time to relax with magnetic field i.e. the relaxaton process become slow. 
\\In Fig.~\ref{fig: 96h_30K} the relaxation behaviour of $\epsilon^\prime$ is shown at 30 K by varying the magnetic field. The difference between dielectric constant with magnetic field and without magnetic field, for both real and imaginary part [$\Delta\epsilon^\prime = \epsilon^\prime (H) - \epsilon^\prime (0)$ similarly, $\Delta\epsilon^{\prime\prime}$] are plotted in Fig.~\ref{fig: 96h_30K} (a) and (b), respectively. It is important to note that the relaxation frequency is not shifted with increasing magnetic field, but the width and height of the peak are increased. It can be speculated that the magnetic field excites more number of charge carriers to relax at the same frequency, and does not change the relaxation time of the charge carrier. 
\\We have measured the dielectric relaxation behaviour of PANI and fMWCNT/PANI composites by applying magnetic field in the perpendicular direction of applied ac electric field and plotted in Fig.~\ref{fig: ac2}. No relaxation peak is observed for PANI at low frequency for applying magnetic field up to 5 T. The MWCNT/PANI exhibits relaxation and the peak intensity increases with increasing magnetic field. However, the relaxation peak frequency does not shift position with magnetic field similar to the applied field in parallel direction. The relaxation behaviour of MWCNT/PANI composite in the both parallel and perpendicular direction imply the inhomogeneity of the composite system is distributed isotropically in all directions e.g. 3D nature of the system. The MWCNT makes network in the PANI and creates interfacial polarization which are isotropically distributed in 3D fashion. We can also observe that another relaxation appears at higher frequency for perpendicular field. This indicates the higher energetic relaxation in the system. 

%\section{Discussion}

Our result is very dissimilar to the doped silicon and RGO composites with different polymers. In doped silicon, Brook et al. observed the anomalous dielectric relaxtion for perpendicular direction however it is usual Debye like for parallel field~\cite{PhysRevB.78.045205Brook}. They have explained that polarized charge carriers feel the Lorentz force due to applied magnetic field in perpendicular direction and modify the polarization. The relaxation peak frequency shifts with increasing magnetic field for a constant temperature. In our case, there is no shift in relaxation peak by increasing magnetic field, and can't be explained completely by considering this model. On the other hand, Parish and Littlewood (PL) proposed that there is possibility to observe the dielectric relaxation [we have mensioned here `relaxation' as it takes place at low frequency though PL used the term `resonance'] in 2D inhomogeneous medium and interface between different conductive materials~\cite{doi:10.1098/rsta.2012.0452M_Parish,PhysRevLett.101.166602Parish} by applying magnetic field without magnetism present in the system. The dielectric relaxation is obeserved due to the mixing of real and imaginary part of the dielectric response occured by inhomogeneity of the medium. The relaxation peak is observed for the condition $\beta\omega\tau$ = 1, where $\beta$ = $\mu$H ($\mu$ is the mobility of charge carrier and H is magnetic field) and $\tau$ = $\rho\epsilon$ ($\rho$ is resistivity). Considering the resistivity variation with temperature is like $\rho$ = $\rho_0$exp($\Delta$/k$_B$T); $\Delta$ is activation energy, the relaxation condition modifies to ln\big($\frac{1}{\beta\omega\epsilon\rho_0}$\big)= $\dfrac{\Delta}{k_BT}$. Assuming no variation of dielectric constant with temperature, modified relaxation condition predicts that the relaxation peak shifts towards higher frequency with increasing temperature for a constant magnetic field, and it should shift towards lower frequency with increasing magnetic field for a constant temperature. 
In case of our sample, the dielectric resonance peak shifts towards lower frequency with increasing temperature for a constant magnetic field which is the opposite behaviour of the prediction. PL model also tells that the dielectric resonance enhances for the current flow in the perpendicular direction of the magnetic field. In the MWCNT/PANI composites, MWCNT is distributed in all the direction (3D nature) and finds the perpendicular current flow at the interface with respect to the magnetic field for which we have observed the dielectric relaxation for both parallel and perpendicular direction.
\\ In summary, we studied the dielectric properties of non-magnetic fMWCNT/PANI composite with varying functionalization, temperature and magnetic field. We have found that the magnetic field has significant effect on the origin of dielectric relaxation. The dielectric relaxation is observed for all the composite at low frequency in the presence of magnetic field. We have tried to explain our results with the essence of 2D PL theory. However, the results are not fully explained by this theory. An extended version of this theory in 3-dimension is required to understand the magnetic field dependence dielectric relaxation. Interestingly, the dielectric relaxation due to inhomogeneity of the system is studied extensively in this fMWCNT/PANI system. We believe this study will help to pave the way of this phenomena in other similar composite systems.

\section{Experimental}

The chemical functionalization of MWCNT is done by using concentrated H$_2$SO$_4$ and HNO$_3$ (vol. ratio 3:1) acid mixtures. The degree of functionalization of MWCNT is varied by changing the duration (hour; denoted as `h'; we have prepared particularly 6 h, 48 h and 96 h) of functionalization~\cite{Maity_2019}. The composites of MWCNT and PANI are prepared by using in-situ polymerization technique which is explained in our previous reports~\cite{Maity_2018}. The samples were made in the form of pallet with radius 10 mm and thickness 1.35 mm. The dielectric measurements are performed keeping the samples in between two silver plates, and put inside the Janis cryostat equipped with superconducting magnet.
The magnetic field is applied both in perpendicular and parallel directions of applied ac electric field by altering the orientation of the sample with respect to the external magnetic field. The dielectric measurements are performed by using Agilent 4294A high precision impedance analyzer.
\\ We have also characterized the samples using SEM, Raman spectroscopy, XPS in our previous reports. It is observed that PANI coated on the surface of MWCNT and distributed randomly in the system. The different functional groups attached  on the surface of MWCNT enhances the polar interaction between fMWCNT and PANI.

\begin{acknowledgement}

Authors thank to Seena Mathew and Husna Jan for helping in measurements. KPM thanks to IISc for reasearch associate fellowship.

\end{acknowledgement}

%\begin{suppinfo}

%A listing of the contents of each file supplied as Supporting Information
%should be included. For instructions on what should be included in the
%Supporting Information as well as how to prepare this material for
%publications, refer to the journal's Instructions for Authors.

%The following files are available free of charge.
%\begin{itemize}
%  \item Filename: brief description
%  \item Filename: brief description
%\end{itemize}

%\end{suppinfo}

%\bibliographystyle{biochem}
%\bibliography{bib}

\end{document}